\documentclass[prd,twocolumn,showpacs,floatfix,superscriptaddress,nofootinbib]{revtex4}
\usepackage[utf8]{inputenc}
\usepackage{graphicx}
\usepackage{epsfig}
\usepackage{bm}
\usepackage{amsfonts}
\usepackage[T1]{fontenc}
\usepackage{amssymb}
\usepackage{float}
\usepackage{amsmath}
\usepackage{dcolumn}
\usepackage{wasysym}
\usepackage{cancel}
\usepackage[colorlinks]{hyperref}
\usepackage[usenames,dvipsnames]{color}
\hypersetup{
     breaklinks=true,
    pdfstartview={FitH},    % fits the width of the page to the window
    colorlinks=true,       % false: boxed links; true: colored links
    linkcolor=blue,          % color of internal links
    citecolor=red,        % color of links to bibliography
    filecolor=magenta,      % color of file links
    urlcolor=blue,           % color of external links
    anchorcolor=green,      % Color for anchor text
    linktocpage=true
}

\def\doi{http://doi.org}

\def\be{\begin{equation*}}
\def\ee{\end{equation*}}

\begin{document}

\title{Anomalous Decay Rate and Greybody Factors for Regular Black Holes with Scalar Hair}

\author{Ram\'on B\'ecar}
\email{rbecar@uct.cl} \affiliation{\small{Departamento de Ciencias Matem\'aticas y F\'{i}sicas, Facultad de Ingenier\'ia, Universidad Cat\'olica de Temuco, Montt 56, Casilla 15-D, Temuco, Chile.}}
\author{P. A. Gonz\'{a}lez}
\email{pablo.gonzalez@udp.cl} \affiliation{Facultad de
Ingenier\'{i}a y Ciencias, Universidad Diego Portales, Avenida Ej\'{e}rcito
Libertador 441, Casilla 298-V, Santiago, Chile.}
\author{Eleftherios Papantonopoulos}
\email{lpapa@central.ntua.gr}
\affiliation{Physics Division, School of Applied Mathematical and Physical Sciences, National Technical University of Athens, 15780 Zografou Campus, Athens, Greece.}
\author{Yerko V\'{a}squez}
\email{yvasquez@userena.cl}
\affiliation{Departamento de F\'{\i}sica, Facultad de Ciencias, Universidad de La Serena,\\
Avenida Cisternas 1200, La Serena, Chile.}

\begin{abstract}
We study the propagation of massive scalar fields in the background of asymptotically flat regular black holes supported by a phantom scalar field with a scalar charge $A$. This parameter regularizes the geometry by removing the central singularity. Focusing on wave dynamics, we analyze scalar perturbations, quasinormal modes, and greybody factors, emphasizing the role of the regularization parameter on the effective potential and the decay properties of the modes. Using WKB methods beyond the eikonal limit, we show that the presence of scalar hair modifies both the oscillation frequencies and damping rates of quasinormal modes. In particular, we demonstrate the occurrence of an anomalous decay rate for massive scalar perturbations, and above a critical field mass, the longest-lived modes correspond to lower angular momentum, in contrast with the massless case. We derive analytical expressions for the critical mass and study its dependence on the scalar charge and overtone number. Furthermore, we apply the Horowitz--Hubeny method to compute the quasinormal frequencies and show that the results obtained from the WKB and Horowitz–Hubeny approaches exhibit excellent agreement in the regime where both methods are valid. In addition, we compute reflection and transmission coefficients and analyze the corresponding greybody factors, clarifying how regularity effects imprint themselves on black-hole scattering properties. Our results show that regular black holes with scalar hair exhibit distinctive dynamical signatures that can be probed through quasinormal ringing and wave propagation.
\end{abstract}

\maketitle

\tableofcontents

\section{Introduction}

The creation of matter at the end of inflation in standard cosmology allows the Universe to enter a deceleration regime and then to return to an accelerated phase  again. The recent observational results indicate that the cosmological evolution at early time, the matter structure formation  of the Universe was governed by a peculiar matter, the dark matter.  Dark matter is a mysterious form of matter. Its properties and characteristics remain largely unknown. Nevertheless, astronomical observations indicate that dark matter accounts for more than 30\% of the total matter–energy content of the Universe.  Dark energy is a peculiar form of energy and may point to still unknown forms of matter-energy in the Universe \cite{,DEDMint1}. The dark energy which was characterized by negative values of the  pressure to density ratio $w$ which could even be  $w < -1$ \cite{steinhardt,tegmark,seljak,hannestad,star03,chandra}. The negative value of $w < -1$ imposes  a constraint on the nature of dark energy and should be parametrized by a {\it phantom field} having negative kinetic energy \cite{sen,gorini,fara05}. However, in this case, a perfect-fluid description of dark energy described by  an imaginary velocity of sound that characterizes the phantom-matter case is  plagued with instabilities at small scales. To avoid this instability, a phantom scalar field may  be regarded as an effective field description following from an underlying theory with positive energies \cite{no03,trod}.

If dark matter can undergo gravitational condensation, star-like objects composed of dark matter may be created in the Universe.  Such dark matter objects could be optically transparent, providing a possible explanation for the central bright spot observed in the Universe. In the same time, the problem of black hole (BH) singularities is a central issue in theoretical physics. It is well known that the cosmic censorship conjecture postulates that singularities  are covered by an one-way event horizon.  In 1968, Bardeen \cite{Regularblackhole1} proposed a novel approach by constructing a BH metric in which the central singularity is removed; such objects are known as regular BHs. Recently, it was claimed that BHs can be supported by astrophysical and cosmological observations as realistic astrophysical BH models \cite{Farrah:2023opk}. In this way non-singular cosmological BH models can couple to the expansion of the Universe, gaining mass proportional to the scale factor.  This claim was based on a recent study of supermassive BHs within elliptical galaxies \cite{akiyama2022first}. This leads to a realistic behavior at infinity of BH models   predicting that the gravitating mass of a BH can increase with the expansion of the Universe  in a manner that depends on the BH interior solution.  Then in \cite{Farrah:2023opk} it was proposed that stellar remnant BHs are the astrophysical origin of dark energy, explaining the onset of accelerating expansion of the Universe.

A gravity theory with a scalar field  minimally coupled to gravity with arbitrary potentials and negative kinetic energy was investigated in \cite{Bronnikov:2005gm}  and local solutions were produced. It was found that regular configurations were formed by the phantom scalar field in  flat, de Sitter (dS) and anti-de Sitter (AdS) asymptotic spacetimes, avoiding the BH central singularity. The motivation of these studies  was to find regular BH solutions with an expanding, asymptotically de Sitter Kantowski-Sachs  cosmology beyond the event horizon.  These geometries corresponds to what are known as black universes. Alternative solutions with a regular center have also been reported in the literature~\cite{dym92,ned01,bdd03}.

A gravitational model in the presence of a self-interacting phantom scalar field  with a scalar charge $A$, minimally coupled to gravity was studied in \cite{Karakasis:2023hni}. If the scalar charge is zero, then the gravitational singularity is covered by a horizon, and then  a normal BH with a constant scalar field is found. However, if $A$ is not zero, then the scalar charge of the phantom scalar field deforms the geometry in such a way that the gravitational singularity is absent and a compact object is generated with a horizon, which is a regular BH. It was found that the  charge of the scalar field is connected to the mass of the BH dressing in this way the BH with secondary hair. 

Quasinormal modes (QNMs) provide a direct link between the perturbative dynamics of compact objects and gravitational-wave observations \cite{Kokkotas:1999bd,Berti:2009kk,Konoplya:2011qq}. In particular, the ringdown phase of a binary merger is governed by a superposition of damped oscillations whose frequencies and decay rates are determined by the underlying spacetime geometry. Therefore, any modification of the effective potential, such as those arising in black-universe configurations \cite{Bronnikov:2005gm}, may lead to observable deviations in the quasinormal spectrum. The linear stability of these configurations was investigated in \cite{Bronnikov:2012ch}, where it was shown that stability is not a generic feature, but rather occurs only within a restricted region of parameter space. In particular, while most configurations are unstable under spherically symmetric perturbations, there exists a special subset of solutions—characterized by specific parameter relations—for which the system remains stable. Outside this regime, the effective potential typically develops negative gaps that can trigger unstable modes or significantly modify the damping times. Even within the stable sector, the modified global structure of these geometries—namely, the presence of a regular interior connected to an expanding region beyond the horizon—affects the shape of the effective potential and, consequently, the associated quasinormal spectrum.

From an observational perspective, these effects would manifest as shifts in the ringdown frequencies and decay rates relative to those predicted for standard black holes. Such deviations could, in principle, be probed through black hole spectroscopy with current and future gravitational-wave detectors \cite{Dreyer:2003bv,Berti:2005ys,Cardoso:2016rao, Isi:2019aib}. However, a quantitative assessment of detectability requires embedding these solutions into realistic astrophysical scenarios and constructing full waveform templates \cite{Buonanno:2000ef,Ajith:2007kx,Husa:2015iqa,Khan:2015jqa}.

When a black hole spacetime is perturbed by a massless scalar field, the longest-lived quasinormal modes are those with higher angular number $\ell$. However,  in the presence of a massive scalar field: above a certain critical value of the scalar field mass, the hierarchy of damping times is inverted, and the longest-lived modes become those with lower angular number. This inversion can be understood from the additional energy stored in the massive perturbation, which reshapes the subleading structure of the effective potential and modifies the decay pattern of the modes. The phenomenon is known to occur in asymptotically flat, asymptotically dS, and asymptotically AdS spacetimes; however, the existence of a critical mass has been established only for asymptotically flat and asymptotically dS geometries, as it does not arise for large or intermediate AdS black holes. The anomalous decay of quasinormal modes has been widely explored for scalar fields~\cite{Konoplya:2006br,Dolan:2007mj,Lagos:2020oek,Aragon:2020tvq,Aragon:2020xtm,Fontana:2020syy,Becar:2023jtd,Becar:2023zbl,Becar:2024agj,Becar:2025niq,Kouniatalis:2025pxs}, for charged scalar perturbations~\cite{Gonzalez:2022upu,Becar:2022wcj}, and for fermionic fields~\cite{Aragon:2020teq}. The effect has also been analyzed in accelerating black holes~\cite{Destounis:2020pjk}, and more recently in wormhole geometries~\cite{Gonzalez:2022ote,Alfaro:2024tdr}, showing that it is a robust feature in a broad class of gravitational backgrounds.

In a recent study~\cite{Gonzalez:2025yjm}, following \cite{Bronnikov:2005gm, Karakasis:2023hni} we considered a phantom scalar field with a scalar charge $A$ in a four-dimensional gravity theory, and analyzed timelike geodesic motion focusing on bound and unbound trajectories and on weak-field observables. In particular, we derived the perihelion precession of bound orbits and used Solar System data to constrain the parameter controlling the regularization of the geometry. These results provide an independent, observationally motivated characterization of the underlying spacetime and delimit the physically relevant region of the parameter space. In the present work, we complement that particle-dynamics perspective with a wave-dynamics analysis by studying the propagation of massive fields, quasinormal ringing, and transmission/reflection probabilities, thereby clarifying how the same regularizing parameter imprints itself both on orbital tests and on the dynamical response of the black hole.

The paper is organized as follows. In Sec.~\ref{setup}, we present the theoretical setup of regular black holes and summarize the main properties of the corresponding spacetime geometry. Sec. \ref{QNM} is devoted to the analysis of scalar field perturbations propagating on this background. In Sec.~\ref{ABQNMS}, we investigate the photon-sphere quasinormal modes and analyze the anomalous behavior of the decay rate of the modes. Sec.~\ref{grey} focuses on the computation of greybody factors. Finally, Sec.~\ref{FR} contains our conclusions.
  
\section{Setup of the theory of Regular black holes}\label{setup}

We consider the action
\begin{equation} S = \int d^4x \sqrt{-g}\left[ \frac{R}{2\kappa} - \frac{1}{2}f(\phi)\nabla_{\mu}\phi\nabla^{\mu}\phi - V(\phi)\right]~,\end{equation}
where $R$~\footnote{Our conventions and definitions throughout this paper are: $(-,+,+,+)$ for the signature of the metric, the Riemann tensor is defined as
$R^\lambda_{\,\,\,\,\mu \nu \sigma} = \partial_\nu \, \Gamma^\lambda_{\,\,\mu\sigma} + \Gamma^\rho_{\,\, \mu\sigma} \, \Gamma^\lambda_{\,\, \rho\nu} - (\nu \leftrightarrow \sigma)$,
and the Ricci tensor and scalar are given by  $R_{\mu\nu} = R^\lambda_{\,\,\,\,\mu \lambda \nu}$ and $R= g^{\mu\nu}\, R_{\mu\nu}$ respectively.} is the Ricci scalar, $\kappa=8\pi G$ with $G=1$, and $\phi$ is a self interacting scalar field. The function $f(\phi)$ determines the nature of the scalar field: canonical if $f>0$, phantom if $f<0$. The Einstein and scalar field equations are
\begin{eqnarray}
&&G_{\mu\nu} =\kappa T_{\mu\nu} ~,\\
&&f(\phi)\Box\phi +\frac{f'(\phi)}{2}\nabla_{\mu}\phi\nabla^{\mu}\phi  = \frac{dV}{d\phi}~,
\end{eqnarray}
with
\begin{equation}
T_{\mu\nu} = f(\phi) \nabla_{\mu}\phi\nabla_{\nu}\phi - \frac{f(\phi)}{2}g_{\mu\nu}\nabla_{\alpha}\phi\nabla^{\alpha}\phi - V(\phi)~.
\end{equation}

We consider a static, spherically symmetric metric
\begin{equation}  \label{ds}
ds^2 = - b(r)dt^2 + \cfrac{1}{b(r)}dr^2 + w(r)^2 d\Omega^2  ~,\end{equation}
where
\begin{equation} d\Omega^{2} = d\theta^2 + \sin^{2}\theta d\varphi^2, \end{equation}
and
\begin{equation}
w(r) = \sqrt{r^2+A^2}~, \label{defor}
\end{equation}
where the parameter $A$ is a length scale.
In \cite{Bronnikov:2005gm} a class of static, spherically symmetric regular black hole solutions with a phantom scalar field was constructed and the associated stress–energy tensor was analyzed, including the evaluation of the corresponding energy conditions.

For a phantom scalar field ($f(\phi)=-1$), one finds the following exact solution:
\begin{eqnarray}\label{bhor}
\notag b(r) &=& c_1 \left(A^2+r^2\right)\\
&&-\frac{c_2 \left(\left(A^2+r^2\right) \tan ^{-1}\left(\frac{r}{A}\right)+A r\right)+2 A r^2}{2 A^3}~,
\end{eqnarray}
and the scalar field and the potential reads
\begin{eqnarray}\label{d4solution}
\phi(r) &=& \frac{1}{2 \sqrt{\pi }}\tan ^{-1}\left(\frac{r}{A}\right)~, \\
V(\phi) &=& \frac{4 A \left(A^2 c_1-1\right) \left(\cos \left(4 \sqrt{\pi } \phi \right)-2\right)}{32 \pi  A^3}\\
\notag && +\frac{c_2 \left(3 \sin \left(4 \sqrt{\pi } \phi \right)-4 \sqrt{\pi } \phi  \left(\cos \left(4 \sqrt{\pi } \phi \right)-2\right)\right)}{32 \pi  A^3}~.
\end{eqnarray}
The asymptotic expansion of the scalar field at spatial infinity yields
\begin{align}
\phi(r \to \infty) = \frac{\sqrt{\pi} A}{4 \, |A| } - \frac{1}{2\sqrt{\pi}}\, \frac{A}{r} + \mathcal O\left(\frac{A^3}{r^3}\right)\, .
\end{align}
From this expression it follows that the parameter $A$ determines the leading $1/r$ fall-off of the scalar field and therefore can be interpreted as a scalar charge.

The large-$r$ expansion of the metric function yields
\begin{eqnarray} 
\notag b(r\to\infty) &\sim& r^2 \left(c_1-\frac{4 A+\pi  c_2}{4 A^3}\right)+\left(A^2 c_1-\frac{\pi  c_2}{4 A}\right)\\
&&+\frac{c_2}{3 r}-\frac{A^2 c_2}{15
   r^3}+\mathcal{O}\left(1/r^5\right)~.
\end{eqnarray}
In this form, the leading behavior of $b(r)$ resembles that of a Schwarzschild-(Anti-)de Sitter spacetime, with an effective cosmological term proportional to $r^2$ and subleading corrections that depend on the scalar charge $A$ and the integration constants. To clarify the asymptotic structure, one can perform a coordinate redefinition $r^2 = R(r)^2-A^2$, in terms of a new radial coordinate $R$, after which the expansion becomes
\begin{eqnarray} 
\notag b(R\to\infty) &\sim& R^2 \left(-\frac{\pi  c_2}{4 A^3}-\frac{1}{A^2}+c_1\right)\\
&&+1+\frac{c_2}{3 R}+\mathcal{O}\left(1/R^3\right)\,.
\end{eqnarray}
In the above expansion there is no deficit angle at spatial infinity, indicating that the geometry does not correspond to a global monopole but rather to a genuine black hole spacetime free of angular deficits. 

To ensure that the spacetime is asymptotically flat, one can fix the integration constants by imposing
\begin{equation} A^2 c_1-\frac{\pi  c_2}{4 A}=1 \to c_1=\frac{4 A+\pi  c_2}{4 A^3}~.\end{equation}
With this choice, the asymptotic expansion of the metric function simplifies to
\begin{equation} b(r) \sim 1+\frac{c_2}{3 r}-\frac{A^2 c_2}{15 r^3}+\frac{A^4 c_2}{35 r^5}+\mathcal{O}\left(1/{r^7}\right)~.
\end{equation}
In this form, the spacetime is manifestly asymptotically flat, approaching Minkowski space at large $r$, and closely resembles the Schwarzschild solution up to corrections sourced by the phantom scalar charge $A$. 
For the present solution, the conserved mass can be computed using the Abbott-Deser prescription for asymptotically flat spacetimes:
\begin{equation}\label{d4mass}
m = \frac{1}{2}\lim_{r\to\infty} r\left(\frac{1}{b(r)} -1\right) = -\frac{c_2}{6}~,
\end{equation}
demonstrating that the scalar field does not directly contribute to the conserved mass. Consequently, the spacetime describes a regular, asymptotically flat black hole carrying a primary phantom scalar hair without altering the global mass of the system.
Using  $m=-c_2/6$, the scalar potential can be written as
\begin{eqnarray} \label{d4pot}
\notag V(\phi) &=& \frac{3m}{16 \pi A^3} \Bigg[- 8 \sqrt{\pi } \phi -3 \sin \left(4 \sqrt{\pi } \phi \right)\Big)\\
&& +2 \pi - \left(\pi -4 \sqrt{\pi } \phi \right) \cos \left(4 \sqrt{\pi } \phi \right)  \Bigg] ~.
\end{eqnarray}
The horizon is determined by the roots of $b(r)=0$. However, this equation cannot be solved analytically in closed form. To understand the existence of a horizon, it is useful to examine the behavior of $b(r)$ near the origin
\begin{equation} b(r\to0)\sim \frac{4 A-6 \pi  m}{4 A}+\frac{6 m r}{A^2}-\frac{3 (\pi  m) r^2}{2 A^3}+\mathcal{O}\left(r^3\right)~,\end{equation}
where the dominant contribution is the constant term $(4A - 6 \pi m)/4A$. The sign of the leading constant term controls the existence of a finite root. For $r>0$, the linear term is positive, so a zero requires $m>2A/3\pi$. Conversely, for $r<0$ the linear term is negative, yielding a finite negative root only if $m<2A/ 3\pi$. At the critical value $m=2A/3\pi$, the constant term vanishes and the horizon is located at $r=0$, coinciding with the minimum of the area function $w(r)$.
The linear stability of these configurations under spherically symmetric perturbations was analyzed in \cite{Bronnikov:2012ch}. Although most configurations are unstable under radial perturbations, a restricted subset of solutions satisfying specific parameter relations remains stable. This occurs precisely when the event horizon coincides with the minimum of the areal radius, that is $m= 2A/3 \pi$.
\newline

\begin{figure}[!h]
	\begin{center}
\includegraphics[width=80mm]{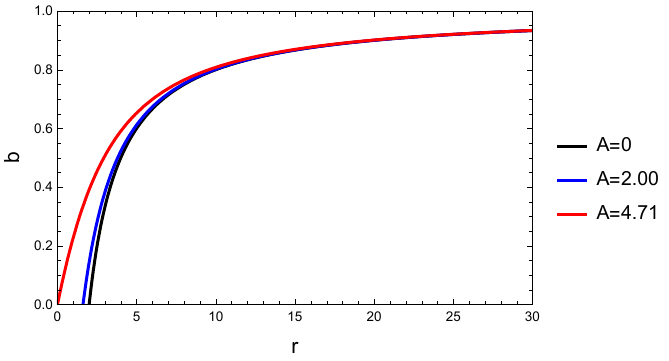}
	\end{center}
	\caption{Plot of the lapse function $b(r)$.
		Here we have used the value $m=1$. The event horizon is at $r_+=2.000$ for $A=0$,  $r_+=1.610$ for $A=2.00$, and $r_+=0$ for $A=4.71$.}
	\label{f1}
\end{figure}

Asymptotically  flat black hole.  Let us consider the asymptotically flat solution, $c_1=\frac{1}{A^2}-\frac{3\pi m}{2A^3}$,  $c_2=-6m$  and $A \leq \frac{3\pi m}{2}$. The lapse function $b(r)$ is given by
\begin{eqnarray}\nonumber
b(r) &\equiv& \frac{3m}{A^2}r-\frac{r^2}{A^2}+\frac{(4A-6\pi m)(r^2+A^2)}{4\,A^{3}}+\\ \label{bflat0}
&+&\frac{3 m(r^2+A^2)}{\,A^{3}}\arctan{r\over A}\,.
\end{eqnarray}
Fig. \ref{f1} shows the radial profile of the lapse function $b(r)$ for different values of the parameter $A$. In the Schwarzschild limit ($A=0$), the lapse vanishes at the horizon, recovering the standard  black hole solution. However, for $A>0$, the family of solutions corresponds to regular black holes, where the central singularity is removed while maintaining asymptotic flatness. Increasing values of $A$ shift the position of the event horizon toward smaller values of the radial coordinate $r$. However, the invariant areal radius of the horizon, $w(r_+)$, actually increases with $A$. At the same time, the spacetime smoothly approaches the flat limit at large distances. At the critical value $A\simeq 4.71$, the lapse vanishes only at the limit $r_+ =0$, at which point the horizon matches the minimum of the area function.

As we discussed in this section, the parameter  A is a phantom scalar hair and when the scalar field backreacts to the metric, A becomes a primary scalar charge. Above all, the most important role of this parameter is that it makes the black hole regular.

Note, however, that the mass parameter $m$, which arises as an integration constant, appears explicitly in the scalar potential (\ref{d4pot}), i.e. in a quantity that defines the theory itself. Consequently, varying the black hole mass would effectively modify the underlying theory rather than simply selecting a different solution within the same model. Therefore, we introduce the constant $c=m/A^3$, which is treated as the parameter entering the scalar potential. With this redefinition, the mass parameter is no longer independent of the scalar charge $A$, and the scalar hair becomes secondary, i.e., it is determined by the black hole parameters rather than introducing an additional independent charge.
With this parameterization, the scalar potential no longer depends on the scalar charge:
\begin{eqnarray}
\notag V(\phi) &=& \frac{3 c}{16 \pi} \Bigg[-8 \sqrt{\pi } \phi -3 \sin \left(4 \sqrt{\pi } \phi \right)\\
&&+ 2 \pi - \left(\pi -4 \sqrt{\pi } \phi \right) \cos \left(4 \sqrt{\pi } \phi \right)  \Bigg] \,,
\end{eqnarray}
and the metric function can be written as
\begin{eqnarray} 
\nonumber b(r) &=& 3 c A r-\frac{r^2}{A^2}+\frac{(4-6\pi c A^2)(r^2+A^2)}{4\,A^2}+\\ \label{bflat0}
&+&3 c (r^2+A^2)\arctan{r\over A}\,.
\end{eqnarray}
Expanding $b(r)$ for small $A/r$ yields
\begin{equation}
b(r) \approx 1 - \frac{2 c A^3}{r} + \frac{2 c A^5}{5 r^3} + \mathcal{O}(A^7)
\end{equation}
and in the limit $A=0$ the metric reduces to Minkowski spacetime.  With this parameterization, the event horizon coincides with the minimum of the areal radius for $c=2/(3 \pi A^2)$. In the following, we study how this parameter influences the QNMs and greybody factors.

\section{Scalar field perturbations}
\label{QNM}

The QNMs of scalar perturbations in the background of the metric (\ref{ds}) for the asymptotically flat black hole defined by $b(r)$ in (\ref{bflat0}), are determined by solving the Klein-Gordon equation for a scalar field in this spacetime 
\begin{equation}
\frac{1}{\sqrt{-g}}\partial _{\mu }\left( \sqrt{-g}g^{\mu \nu }\partial_{\nu } \varphi \right) = \bar{m}^{2}\varphi \,,  \label{KGNM}
\end{equation}%
with suitable boundary conditions for a black hole geometry. In the above expression $\bar{m}$ is the mass
of the scalar field $\varphi $. Now, by means of the following ansatz
\begin{equation}
\varphi =e^{-i\omega t} R(r) Y(\Omega) \,,\label{wave}
\end{equation}%
the Klein-Gordon equation reduces to
\begin{eqnarray}
\notag && b(r)\partial ^2_rR(r)+\left(\frac{2w'(r)b(r)}{w(r)}+b'(r)\right)\partial_r R(r)+ \\
&& \left( \frac{\omega^2}{b(r)}+\frac{\kappa}{w^2(r)}-\bar{m}^2\right)R(r)=0\,,
\label{radial}
\end{eqnarray}
where we defined $\kappa=-\ell (\ell+1)$, with $\ell=0,1,2,...$, which represents the eigenvalue of the Laplacian on the two-sphere and $\ell$ is the multipole number.
Now, by using the tortoise coordinate $r^*$ given by
$dr^*=\frac{dr}{b(r)}$, and then defining $R(r)=\frac{F(r)}{w(r)}$ the Klein-Gordon equation can be written as an one-dimensional Schr\"{o}dinger-like equation
 \begin{equation}\label{ggg}
 \frac{d^{2}F(r^*)}{dr^{*2}}-V_{\text{eff}}(r)F(r^*)=-\omega^{2}F(r^*)\,,
 \end{equation}
 with an effective potential $V_{\text{eff}}(r)$, which  parametrically thought,  $V_{\text{eff}}(r^*)$, is given by
  \begin{equation}\label{pot}
 V_{\text{eff}}(r)=b(r) \left(\bar{m}^2 -\frac{\kappa}{w(r)^2} +  \frac{(b(r)w^{\prime}(r))^\prime}{w(r)} \right)~.
 \end{equation}

\begin{figure}[!h]
	\begin{center}
\includegraphics[width=80mm]{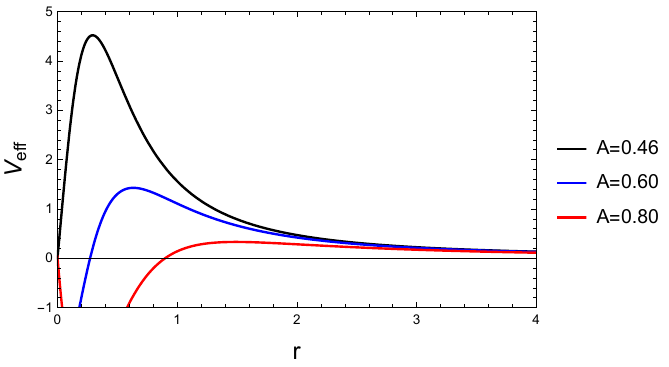}
	\end{center}
	\caption{Plot of the effective potential as a  function of $r$. Here, we have used the value $c=1$, $\bar{m}=0.1$ and $\ell=1$. The event horizon is at $r_+=0$ for $A=0.46$, $r_+=0.273$ for $A=0.60$, and $r_+=0.901$ for $A=0.80$.}
	\label{fpot}
\end{figure}

Fig.~\ref{fpot} presents the effective potential $V_{\mathrm{eff}}(r)$ for a massive scalar field ($\bar{m}=0.1$) and different values of the deformation parameter $A$. 
The potential displaying a single peak that governs the propagation and trapping of scalar perturbations. 
As $A$ increases, both the height and width of the potential barrier decrease and its maximum shifts slightly toward larger radii. 
This behavior indicates that the deformation weakens the near-horizon confinement and displaces the dominant interaction region outward, while the spacetime remains asymptotically flat.

\section{Photon Sphere Modes}
\label{ABQNMS}

In this section, we use the WKB approximation and the Horowitz–Hubeny method to compute the quasinormal frequencies (QNFs) associated with photon-sphere modes.
\newline

\subsection{WKB approximation}
In this section, we employ the WKB approximation method, originally proposed by Mashhoon~\cite{Mashhoon} and subsequently developed by Schutz and Iyer~\cite{Schutz:1985km}. The third–order formulation introduced by Iyer and Will~\cite{Iyer:1986np} was later extended to sixth order \cite{Konoplya:2003ii}, and more recently to thirteenth order \cite{Matyjasek:2017psv}; see also \cite{Konoplya:2019hlu} for a comprehensive overview. This semi-analytic technique has been successfully applied to compute QNFs of both asymptotically flat and asymptotically de Sitter black holes. The WKB method is particularly well suited for effective potentials of barrier type, whose asymptotic behavior approaches constant values at the horizon and at spatial infinity~\cite{Konoplya:2011qq}. However, it should be emphasized that this method captures only the modes associated with the photon sphere. The QNMs are determined by the behavior of the effective potential in the vicinity of its maximum $V(r^{*}_{\text{max}})$, around which the potential can be expanded in a Taylor series of the form

\begin{equation} \label{expansion}
V(r^*)= V(r^*_{max})+ \sum_{k=2}^{k=\infty} \frac{V^{(k)}}{k!} (r^*-r^*_{max})^{k} \,,
\end{equation}
where
\begin{equation}
V^{(k)}= \frac{d^{k}}{d r^{*k}}V(r^*)|_{r^*=r^*_{max}}\,,
\label{eq:derivadas}
\end{equation}
corresponds to the $k$-th derivative of the potential with respect to $r^*$, evaluated at the position of the maximum of the potential, $r^*_{max}$. Using the WKB approximation up to third order beyond the eikonal limit, the QNFs are given by the following expression \cite{Iyer:1986np, Hatsuda:2019eoj}
\begin{eqnarray}
\omega^2 &=& V(r^*_{max})  -2 i U \,,
\end{eqnarray}
where
\begin{widetext}
\begin{eqnarray}
\notag U &=&  N\sqrt{-V^{(2)}/2}+\frac{i}{64} \left[ -\frac{1}{9}\frac{V^{(3)2}}{V^{(2)2}} (7+60N^2)+\frac{V^{(4)}}{V^{(2)}}(1+4 N^2) \right] +\frac{N}{2^{3/2} 288} \Bigg[ \frac{5}{24} \frac{V^{(3)4}}{(-V^{(2)})^{9/2}} (77+188N^2) + \\
\notag && \frac{3}{4} \frac{V^{(3)2} V^{(4)}}{(-V^{(2)})^{7/2}}(51+100N^2) +\frac{1}{8} \frac{V^{(4)2}}{(-
V^{(2)})^{5/2}}(67+68 N^2)+\frac{V^{(3)}V^{(5)}}{(-V^{(2)})^{5/2}}(19+28N^2)+\frac{V^{(6)}}{(-V^{(2)})^{3/2}} (5+4N^2)  \Bigg]\,,
\end{eqnarray}
\end{widetext}
and $N=n+1/2$, with $n=0,1,2,\dots$, is the overtone number.

Defining $L^2= \ell (\ell+1)$, we find that for large values of $L$, the maximum of the potential is approximately at
\begin{equation}
    r_{max} \approx r_{0}+\frac{r_{1}}{L^{2}}\,,
\end{equation}
where
\begin{equation}
r_{0}=3 c A^3 \,,
\end{equation}
%\begin{widetext}
\begin{eqnarray}
\notag r_{1}&=& \frac{3}{2} c A^3 \xi \eta \Big[  -27 \pi  A^6 c^3+18 A^4 c^2-3 \pi  A^2 c  \\
&+& 1 + A^2 \xi \left(6 c \arctan \left(3 A^2 c\right)+\bar{m}^2\right)  \Big] \,,  
\end{eqnarray}
with $\eta=2-3 \pi  c A^2 + 6 c A^2  \arctan \left( 3 c A^2\right)$, and $\xi=1+ 9 c^2 A^4$.
So, the maximum of the potential is
\begin{equation}
V(r^{*}_{max})\approx V_{0}L^{2}+V_{1} \,,
\end{equation}
where
\begin{equation}
V_{0}= \frac{ \eta}{2 A^2}\,,
\end{equation}
\begin{equation}    
 V_{1}=\frac{\eta}{4 A^2} \Bigg[2 \bar{m}^2 A^2 \xi+\left(1+18 c^2 A^4\right) \eta \Bigg]\,,  
\end{equation}
while the higher order derivatives $V^{(k)}(r^{*}_{max})$ for $k=2,...,6$, can be expressed in the following abbreviated manner
\begin{align}
    V^{(2)}(r^{*}_{max})&\approx V_{0}^{(2)}L^{2}+V_{1}^{(2)} \,, \\
    V^{(3)}(r^{*}_{max})&\approx V_{0}^{(3)}L^{2} \,, \\
    V^{(4)}(r^{*}_{max})&\approx V_{0}^{(4)}L^{2} \,, \\
    V^{(5)}(r^{*}_{max})&\approx V_{0}^{(5)}L^{2} \,, \\
    V^{(6)}(r^{*}_{max})&\approx V_{0}^{(6)}L^{2}~.
\end{align}
where
\begin{equation}
V_0^{(2)} = -\frac{ \eta ^2}{2 A^4}\,,
\end{equation}

\begin{eqnarray}
\notag V_1^{(2)} &=& -\frac{\eta^2 }{4 A^4} \Bigg[ \eta \Big( 1215 \pi  A^{10} c^5-810 A^8 c^4+162 \pi  A^6 c^3 \\
\notag && -54 A^4 c^2+3 \pi  A^2 c+2 + \big(-2430 A^{10} c^5\\
\notag && -324 A^6 c^3 -6 A^2 c\big) \arctan \left(3 A^2 c\right)  \Big) + 3 c A^4 \bar{m}^2 \xi   \\
\notag &&  \times\bigg( 15 A^2 c \left(3 \pi  A^2 c-2\right)\\
&& -2 \left(45 A^4 c^2+1\right) \arctan\left(3 A^2 c\right) +\pi \bigg) \Bigg] \,, 
\end{eqnarray}
\begin{equation}
 V_0^{(3)} =  \frac{3 c \eta^3}{2 A^3} \,, 
\end{equation}
\begin{equation}
 V_0^{(4)} = \frac{3 \left( 2 -\pi c A^2+2 c A^2 \arctan (3 c A^2) \right) \eta^3}{2 A^6} \,, \end{equation}
\begin{equation}
V_0^{(5)} = -\frac{3 c \eta^4 \left( - 3 \pi c A^2 +17 + 6 c A^2 \arctan(3 c A^2)\right)}{2 A^5} \,.
\end{equation}

\begin{eqnarray}
\notag V_0^{(6)} &=& -\frac{9 \eta ^4}{2 A^8} \Bigg[ A^2 c \left(A^2 c \left(45 \pi  A^2 c+\pi ^2-30\right)-10 \pi \right)+ \\
\notag && 2 A^2 c \arctan \left(3 A^2 c\right) \big(-45 A^4 c^2-2 \pi  A^2 c+ \\
&& 2 A^2 c \arctan \left(3 A^2 c\right)+10\big)+10\Bigg]\,.  
\end{eqnarray}

Moreover, our interest is to evaluate the QNFs for large values of $L$, so we expand the frequencies as power series in $L$. 
It is important to keep in mind that in the eikonal limit, the leading term is linear in $L$.
%, and for $\tilde{\alpha}=0$, we should recover the frequencies of the Schwarzschild de Sitter black hole.
Next, we consider the following expression in powers of $L$
\begin{equation}
\label{omegawkb}
\omega=\omega_{1m}L+\omega_{0}+\omega_{1}L^{-1}+\omega_{2}L^{-2} + \mathcal{O}(L^{-3})\,,
\end{equation}
where
\begin{equation}
\notag \omega_{1m}= \frac{1}{3 \sqrt{3} A^3 c} -\frac{1}{90 \sqrt{3} A^7 c^3} + \frac{79}{113400 \sqrt{3} A^{11} c^5} + \dots \,,
\end{equation}

\begin{equation}
\notag \omega_{0}= -\frac{i (2 n+1)}{6 \sqrt{3} A^3 c}+\frac{i (2 n+1)}{180 \sqrt{3} A^7 c^3} - \frac{79 i (2 n+1)}{226800 \sqrt{3} A^{11} c^5} + \dots \,, 
\end{equation}

\begin{equation}
\notag \omega_{1} = \frac{1}{2} \sqrt{3} A^3 c \bar{m}^2 +\frac{7 \bar{m}^2}{60 \sqrt{3} A c}+\frac{17-15 n (n+1)}{324 \sqrt{3} A^3 c} + \dots  \,, 
\end{equation}

%\begin{widetext}
\begin{eqnarray}
\label{omega2}
\notag \omega_2 &=& \frac{5 i A^3 c \bar{m}^2 (2 n+1)}{4 \sqrt{3}}-\frac{i (2 n+1) (235 n (n+1)+137)}{23328 \sqrt{3} A^3 c} \\
\notag && +\frac{i \bar{m}^2 (2 n+1)}{24 \sqrt{3} A c} + \dots   \,,
\end{eqnarray}
 by performing a series expansion in $A/ r_+ $. The ellipsis denote terms of  higher order in $A/r_+$. In Appendix \ref{appendixA}, we show the full expression obtained without approximations.

The term proportional to $1/L^2$ in Eq. (\ref{omegawkb}) vanishes at the value of the critical mass $\bar{m}_c$, which is given by
\begin{eqnarray}
\label{mc}
\notag \bar{m}_c &=& \frac{1}{A^3 c}  \sqrt{\frac{137 + 235 n (n+1) }{29160}}+  \\
&& \frac{151 -235  n (n+1)}{1620 A^7 c^3  \sqrt{10 (137+235 n (n+1))}} + \dots \,.
\end{eqnarray} 
The first term in the expansion resembles the critical scalar field mass of the Schwarzschild case reported in \cite{Lagos:2020oek}.  The scalar charge $A$ decreases the critical mass. The full expression for the critical mass is given in the appendix \ref{appendixA}. Fig. \ref{FMC} shows the dependence of the critical mass $\bar{m}_{c}$ on the deformation parameter $A$ for the overtone numbers $n=0,1,2$. The critical mass shows a monotonically decrease as the parameter $A$ increases for all values of $n$, and tends to zero for high values of $A$, in agreement with (\ref{mc}). Furthermore, for any fixed value of $A$, the critical mass is an increasing function of the overtone number $n$.
\begin{figure}[H]
\begin{center}
\includegraphics[width=0.4\textwidth]{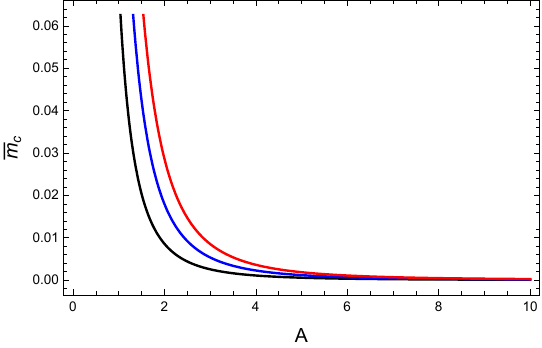}
\end{center}
\caption{The behavior of $\bar{m}_c$ as a function of $A$ for the overtone number $n=0$ (black curve), $n=1$ (blue curve) and $n=2$ (red curve) with $c=1$.} 
\label{FMC}
\end{figure}

\begin{figure}[H]
\begin{center}
\includegraphics[width=0.4\textwidth]{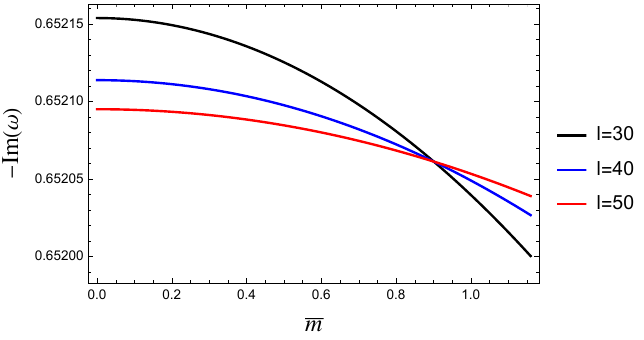}
\includegraphics[width=0.4\textwidth]{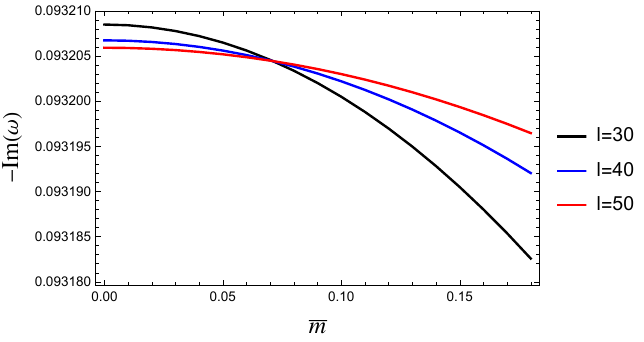}
\includegraphics[width=0.4\textwidth]{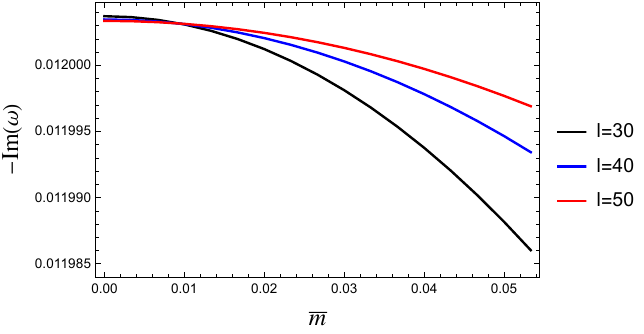}
\end{center}
\caption{The behavior of $-Im(\omega)$ for the fundamental mode ($n=0$) as a function of the scalar field mass $\bar{m}$ for different values of the angular number $\ell=30$ (black curve), $\ell=40$ (blue curve), $\ell=50$ (red curve), with $c=1$, $A=0.4606$ (top panel), $A=1$ (central panel), and $A=2$ (bottom panel) using the 6th order WKB method with Pad\'e approximants. Here, the WKB method yields critical masses of $\bar{m}_{c}\approx 0.901$, $0.071$ and $0.009$, respectively, via Eq. (\ref{mass}).}
\label{AR}
\end{figure}

Now, to illustrate the anomalous behavior, we plot in Fig. \ref{AR} the behavior of $-Im(\omega)$ as a function of $\bar{m}$, using the 6th-order WKB method with Pad\'e approximants. We observe an anomalous decay rate for: $\bar{m}<\bar{m}_c$, the longest-lived modes correspond to the highest angular number $\ell$, while for $\bar{m}>\bar{m}_c$, the longest-lived modes correspond to the lowest angular number.

\begin{figure}[h]
\begin{center}
\includegraphics[width=0.4\textwidth]{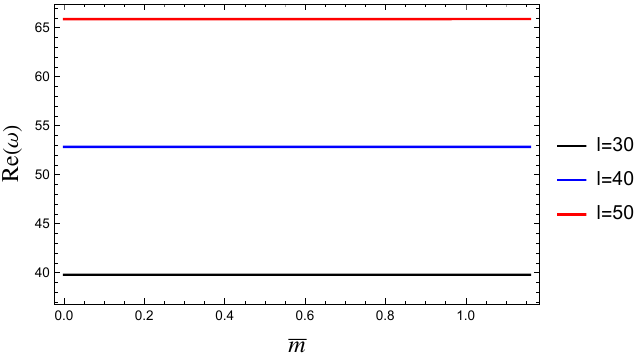}
\includegraphics[width=0.4\textwidth]{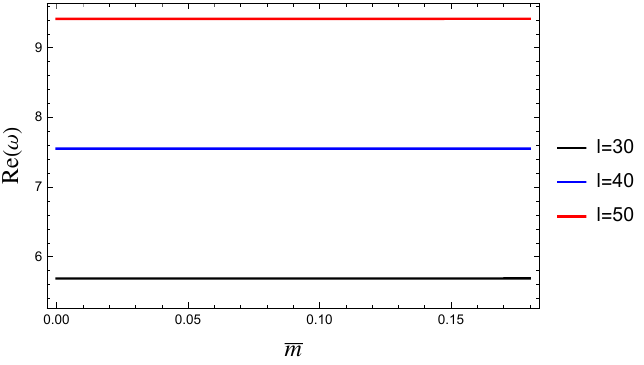}
\includegraphics[width=0.4\textwidth]{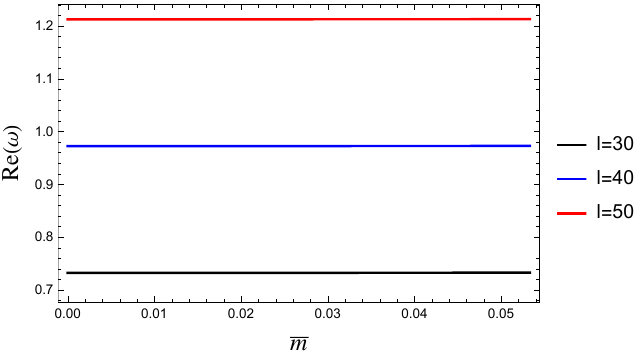}
\end{center}
\caption{The behavior of $Re(\omega)$ for the fundamental mode ($n=0$) as a function of the scalar field mass $\bar{m}$ for different values of the angular number $\ell=30$ (black curve), $\ell=40$ (blue curve), $\ell=50$ (red curve), with $c=1$, $A=0.4606$ (top panel), $A=1$ (central panel), and $A=2$ (bottom panel) using the 6th order WKB method with Pad\'e approximants.}
\label{ARR}
\end{figure}

The behavior shown in Fig.~\ref{ARR} highlights an important contrast between the real and imaginary parts of the quasinormal spectrum. While the imaginary part exhibits a pronounced sensitivity to the scalar field mass, leading to the anomalous decay-rate phenomenon discussed previously, the real part of the frequency remains largely insensitive to $\bar m$. This indicates that the oscillation timescale of the perturbations is predominantly controlled by the geometry near the photon sphere, rather than by the mass of the scalar field.

Moreover, the monotonic decrease of $\mathrm{Re}(\omega)$ with increasing scalar charge $A$ reflects the weakening of the effective potential barrier induced by the regularizing parameter. As $A$ increases, the height of the potential maximum decreases and its location is shifted outward, leading to lower characteristic oscillation frequencies. Nevertheless, the qualitative structure of the spectrum is preserved: the hierarchy in $\ell$ remains unchanged and no inversion analogous to the anomalous behavior observed in the damping rates occurs. 

The origin of the anomalous decay rate can be understood within the WKB framework by expressing the quasinormal spectrum in terms of the effective width of the potential barrier. The QNMs are determined by the behavior of the effective potential near its maximum, where the WKB condition relates the complex frequency to local properties of the barrier.

Rather than parameterizing the barrier in terms of its curvature, it is convenient to characterize it by its width $\Delta r^*$, which measures the extent of the classically forbidden region around the peak. Physically, $\Delta r^*$ controls the trapping efficiency of the potential: wider barriers suppress tunneling and lead to longer-lived modes, whereas narrower barriers enhance leakage and increase the decay rate.

For massive scalar fields, the mass term modifies both the asymptotic structure and the shape of the effective potential, thereby altering the width of the barrier and the system’s ability to store energy. For sufficiently small field masses, increasing the multipole number $\ell$ broadens the effective barrier (larger $\Delta r^*$), enlarging the trapping region and suppressing tunneling. This results in longer-lived modes and gives rise to the anomalous decay behavior. In contrast, above a critical mass scale, the structure of the potential changes such that the barrier effectively narrows as $\ell$ increases. This reduces $\Delta r^*$, enhances tunneling, and leads to faster decay. In this picture, the anomalous decay rate reflects a transition in the effective width of the potential barrier.

To make this relation explicit, we define the width of the effective potential as the interval over which the potential decreases from its maximum value $V(r_{\text{max}}^*)$ to a fraction $\epsilon V$. Expanding the potential to quadratic order around the maximum, one obtains
\begin{equation}
\Delta r^* \approx \sqrt{2(1- \epsilon) \frac{V(r_{\text{max}}^*)}{-V^{(2)}(r_{\text{max}}^*)}} \,.
\end{equation}

Using the WKB expression for the QNFs, where the imaginary part is controlled by the second derivative of the potential at the peak, one finds
\begin{equation}
Im(\omega) \approx -\frac{1}{2 \sqrt{2}} \sqrt{\frac{-V^{(2)}(r_{\text{max}}^*)}{V(r_{\text{max}}^*)}} + \mathcal{O}(1/L^2) \,.
\end{equation}
Substituting the above definition of the width, this can be rewritten as
\begin{equation}
Im(\omega) \approx -\frac{1}{2} \frac{\sqrt{1-\epsilon}}{\Delta r^*} + \mathcal{O}(1/L^2) \,,
\end{equation}
showing explicitly that, in the eikonal limit, the decay rate is inversely proportional to the width of the effective potential barrier \cite{Lagos:2020oek,Kouniatalis:2025pxs}. 
%[27], [31].

\begin{figure}[H]
\begin{center}
\includegraphics[width=0.4\textwidth]{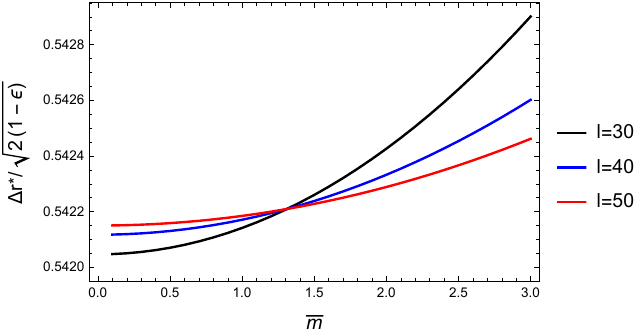}
\end{center}
\caption{Effective width of the potential barrier, $\Delta r^*/ \sqrt{2(1- \epsilon)}$, as a function of the scalar field mass for different values of the multipole number $\ell = 30$ (black curve), $\ell = 40$ (blue curve), $\ell = 50$ (red curve), with $c=1$ and $A=0.4606$.}
\label{WWW}
\end{figure}

In Fig. \ref{WWW} we plot the rescaled width $\Delta r^*/ \sqrt{2(1- \epsilon)}$ as a function of the scalar field mass for different values of the multipole number. For masses below a given threshold, the width increases with $\ell$, whereas for larger masses this behavior is reversed. Although this inversion does not occur exactly at the critical mass, it closely tracks the qualitative change observed in $Im(\omega)$, confirming that the anomalous decay is directly tied to the transition in the effective width of the potential barrier.

\subsection{Horowitz--Hubeny method}

Now, we apply a numerical procedure based on the Horowitz--Hubeny method (HHM) to compute the QNFs \cite{Horowitz:1999jd}. This method was originally developed for perturbations of asymptotically anti-de Sitter (AdS) black holes in the context of holography, where the quasinormal spectrum is related to the relaxation of perturbations in the dual field theory. It is particularly suitable for AdS spacetimes because the perturbation equation has regular singular points at the horizon and at infinity, allowing the solution to be constructed as a power series. The method consists of expanding the radial solution around the event horizon and imposing the boundary condition at spatial infinity, which selects a discrete set of complex frequencies corresponding to the QNMs. Here, we show that the standard HHM can also be applied to asymptotically flat black holes.

In this section we consider the approximation 
\begin{equation}
A/r << 1  \,,  \quad  r_+ < r < \infty  \,,
\end{equation}
which implies $A/r_+ <<1$. Under this assumption,
\begin{equation} \label{primerorden}
b(r) \approx 1 - \frac{2 c A^3}{r} \,,
\end{equation}
and $w(r)^2 = r^2 +A^2$. The event horizon is located at $r_+ \approx 2 c A^3$, therefore
\begin{equation}
\frac{A}{r_+} = \frac{1}{2 c A^2} \,,
\end{equation}
 so the approximation is valid in the regime $c A^2 >> 1$.

To solve the radial equation (\ref{radial}), we impose the boundary conditions satisfied by the radial function $R(r)$. Near the horizon, only ingoing waves are allowed, so
\begin{equation}
 R(r) \sim (r-r_+)^{- i \omega r_+}   \left[  1 + \mathcal{O}((r-r_+))  \right]   \,.
 \end{equation}
 At spatial infinity, we impose outgoing-wave boundary conditions
 \begin{equation}
 R(r) \sim r^{-1 +\frac{i r_+ (2 \omega^2-\bar{m}^2)}{2\sqrt{\omega^2-\bar{m}^2}}} e^{i \sqrt{\omega^2 - \bar{m}^2} r } r \left[  \frac{1}{r} + \mathcal{O}(1/r^2)  \right]\,.
 \end{equation}
Note that a factor $r$ has been deliberately kept outside the brackets to ensure that the quantity within the brackets vanishes as $r \rightarrow \infty$.
Accordingly, we consider the ansantz
\begin{equation}
R(r) = (r-r_+)^{- i \omega r_+} r^{i \omega r_+ -1 +\frac{i r_+ (2 \omega^2-\bar{m}^2)}{2\sqrt{\omega^2-\bar{m}^2}}}  e^{i \sqrt{\omega^2-\bar{m}^2}r} r \psi(r) \,,
\end{equation}
where the function $\psi(r)$ is regular at the horizon and decays as $1/r$ at spatial infinity, thereby satisfying the boundary condition $\psi(r) \rightarrow 0$ as $r \rightarrow \infty$ required by the standard HHM.
Substituting this expression into the radial equation yields a differential equation for $\psi(r)$. We then introduce the compact coordinate $x= 1/r$, for which the horizon is mapped to $x_+ = 1 / r_+$ and infinity to $x = 0$. The radial equation becomes
\begin{equation} \label{radialeq}
s(x) \psi''(x) + \frac{t(x)}{(x-x_+)} \psi'(x) + \frac{u(x)}{(x-x_+)^2} \psi(x) = 0 \,,
\end{equation}
where $s(x)$, $t(x)$, and $u(x)$ are fifth-degree polynomials (see Appendix B). The point $x=x_+$ is a regular singular point.

To compute the quasinormal frequencies, we expand the solution as a power series around the horizon
\begin{equation} \label{exp}
\psi(x) = (x-x_+)^a \sum_{n=0}^{\infty} a_n(x-x_+)^n \,.
\end{equation}
From the indicial equation we obtain
\begin{equation}
 a=0 \,,  \quad   a= 2 i \omega / x_+ \,.
 \end{equation}

 The ingoing boundary condition selects $a=0$. Truncating the series at order $N$ and substituting into the radial equation yields the recurrence relation
\begin{equation}
a_n = - \frac{1}{P_n} \sum_{k=0}^{n-1} \left[  k(k-1)s_{n-k} + k t_{n-k} +u_{n-k} \right] a_k\,,
\end{equation}
with
\begin{eqnarray}
\notag P_n &=& n(n-1)s_0 + n t_0 \\
&=& 4 n x_+ (1+ x_+^2 A^2) (\omega^2-\bar{m}^2)(n x_+ -2 i \omega) \,.
\end{eqnarray}
Since the equation has no poles in $0 < x < x_+$, the quasinormal mode condition is obtained by imposing the boundary condition $\psi(x) \rightarrow 0$ as $x \rightarrow 0$, which leads to the algebraic equation
\begin{equation} \label{serie}
\sum_{n=0}^N a_n (-x_+)^n=0 \,.
\end{equation}
This condition is satisfied only for discrete complex values of $\omega$, corresponding to the quasinormal spectrum.

To determine the zeros of the truncated series equation (\ref{serie}), we employ the M\"uller method \cite{EMuller}, which is a three-point iterative root-finding algorithm particularly suitable for complex roots. The method generalizes the secant method by constructing a quadratic interpolating polynomial through three successive approximations of the function and taking one of its roots as the next iterate. Because it does not require derivatives and naturally handles complex numbers, it is well suited for quasinormal-mode calculations, where the frequency $\omega$ is complex. For sufficiently smooth functions and initial guesses close to the root, the method converges with order approximately 1.84, which is faster than the secant method and typically robust in complex root searches \cite{BurdenFaires}. In practice, we evaluate the truncated series equation for complex values of $\omega$ and iterate until the relative change in the frequency satisfies the chosen numerical tolerance.

In Table I we present the QNFs of a massless scalar field obtained using the HHM and the sixth-order WKB approximation with Pad\'e approximants for $c=1$. The results are reported for several values of the multipole number $\ell$, and for $A=3$ and $A=5$, considering the metric function (\ref{primerorden}). Furthermore, we show the relative error of the real and imaginary parts of the values obtained with the WKB method with respect to the values obtained with the HHM, which is defined by
\begin{eqnarray}
\epsilon_{Re(\omega)} &=& \frac{|Re(\omega_{WKB})-Re(\omega_{HHM})|}{|Re(\omega_{HHM}))|} \cdot 100 \%  \,,   \\
\epsilon_{Im(\omega)} &=& \frac{|Im(\omega_{WKB})-Im(\omega_{HHM})|}{|Im(\omega_{HHM}))|} \cdot 100 \%   \,.
\end{eqnarray}

We find excellent agreement between the two approaches within their common range of applicability. In particular, it is well known that the WKB approximation with Pad\'e improvements yields reliable results when the multipole number $\ell$ is greater than the overtone number $n$, with the accuracy generally improving as $\ell$ increases relative to $n$; for higher overtones with $n \gtrsim \ell$ the WKB results tend to lose precision.
\newline

The stability of the background geometry was analyzed in Ref.~\cite{Bronnikov:2012ch} within the linear approximation, where it was shown that these configurations are generically unstable under spherically symmetric perturbations, except for a special class of solutions in which the event horizon coincides with the minimum of the areal radius; the analysis includes both axial perturbations and the monopole sector of polar perturbations.

In the present work, we instead focus on the propagation of a test scalar field on these fixed backgrounds. Our analysis encompasses both the special class of configurations identified in Ref.~\cite{Bronnikov:2012ch} as stable, as well as cases in which the minimum of the areal function lies inside the horizon. Within this probe approximation, we find no evidence of linear instabilities in the scalar sector for any configuration considered, including those that are unstable under gravitational perturbations. This indicates that, at the level of test-field dynamics, the spacetimes support well-defined quasinormal ringing and late-time behavior without the presence of growing modes.

It is worth emphasizing that, when the horizon coincides with the minimum of the areal function--which occurs for $c= 2/(3 \pi A^2)$--Eqs.~(\ref{A1}) and (\ref{A2}) yield in the eikonal limit
\begin{equation}
\omega = \frac{1}{A} \sqrt{\frac{2}{\pi} \arctan \left( \frac{2}{\pi} \right)} \left( \ell + \frac{1}{2} - i \left( n + \frac{1}{2} \right) \right) + \mathcal{O}(1/\ell),
\end{equation}
in exact agreement with the asymptotic expression reported in Ref.~\cite{Bronnikov:2012ch} for axial gravitational perturbations in this case.

This agreement reflects a more general property of the eikonal regime ($\ell \gg 1$), where the effective potentials governing different types of perturbations share a universal leading-order form. In this limit, the quasinormal spectrum is determined by the properties of unstable null geodesics of the background spacetime, and therefore depends primarily on the geometry rather than on the spin of the perturbing field \cite{Cardoso:2008bp}. This explains why the scalar field QNFs obtained here coincide with those of gravitational perturbations in the same regime.

\begin{table} \label{quasi}
\centering
\caption{The fundamental ($n=0$) QNFs for several values of the angular momentum $\ell$ of a massless scalar field for the asymptotically flat regular black holes with $c=1$, are calculated using the HHM and the sixth-order WKB method with Pad\'e approximants. The QNFs obtained via the HHM have been calculated with nine decimal places of accuracy.}
\footnotesize
\setlength{\tabcolsep}{2pt}
\renewcommand{\arraystretch}{1.05}
\resizebox{\columnwidth}{!}{%
\begin{tabular}{|c|c|c|c|c|}  \hline
\multicolumn{5}{|c|}{$A=3$}  \\  \hline
$\ell$ &  HHM & WKB & $\epsilon_{Re(\tilde{\omega})}(\%)$ & $\epsilon_{Im(\tilde{\omega})}(\%)$ \\
\hline
0  &  0.004086511 - 0.003884059 i & 0.004087112 - 0.003732616 i& 0.015 & 3.9  \\
1  &  0.010841490 - 0.003616272 i &  0.010840493 - 0.003620066 i & 0.009 & 0.105  \\
 2  &  0.017900157 - 0.003582881 i  &  0.017900083 - 0.003583152 i & 0.0004 & 0.0076 \\
3   &  0.024996222 - 0.0035732690 i  &  0.024996214 - 0.003573305 i    &  $3.2\cdot 10^{-5}$  & 0.001 \\
5  &  0.039217876 - 0.003567225 i   &  0.039217875 - 0.003567227 i  & $2.5\cdot 10^{-6}$ & $5.6\cdot 10^{-5}$ \\
10  &   0.074812377 - 0.003564217 i  &  0.074812376 - 0.003564217 i & $1.3\cdot 10^{-6}$ &  0 \\
15  &   0.110419796 - 0.003563600 i &   0.110419796 - 0.003563600 i  & 0 & 0  \\
\hline
\multicolumn{5}{|c|}{$A=5$} \\ \hline
0  & 0.000883516 - 0.000839139 i  &  0.000883635 - 0.000806476 i & 0.013 & 3.9 \\
  1  &  0.002343265 - 0.000781259 i  &  0.002343052 - 0.000782072 i & 0.009 &  0.104 \\
  2  &  0.003868799 - 0.000774048 i &  0.003868782 - 0.000774107 i & 0.0004  & 0.0076 \\
  3   &  0.005402444 - 0.000771975 i &  0.005402442 -  0.000771983 i  & $3.7\cdot 10^{-5}$ & 0.001  \\
   5  &   0.008476138 - 0.000770672 i  &  0.008476137 - 0.000770672 i &  $1.2\cdot 10^{-5}$   & 0  \\
      10  &  0.016169124 - 0.000770024  i &  0.016169123 - 0.000770023 i   & $6.2\cdot 10^{-6}$ & 0.00013  \\
  15  & 0.023864910 - 0.000769890 i &  0.023864909 - 0.000769890 i & $4.2\cdot 10^{-6}$ & 0   \\
\hline
\end{tabular}}
\label{TableI}
\end{table}

\section{Greybody Factors}
\label{grey}

For frequencies satisfying $\omega^{2} \approx V_{0}$, we make use of the WKB approximation beyond the eikonal limit. In particular, we employ the first-order WKB formula developed by Schutz and Will (see Ref.~[36]) for black-hole scattering, which yields
\begin{equation}
R=\left(1+e^{-2 i\pi \left(\nu + \frac{1}{2}\right)}\right)^{-1/2}, 
\qquad \omega^{2}\simeq V_{0},
\label{eq:WKB_R}
\end{equation}
and
\begin{equation}
|T(\omega)|^{2} = 1-|R(\omega)|^{2},
\end{equation}
where $T(\omega)$ denotes the transmission coefficient. The quantity $\nu$ satisfies
\begin{equation}
\nu+\frac{1}{2}
 = i\,\frac{\omega^{2}-V_{0}}{\sqrt{-2 V_{0}''}}
 + \Lambda_{2}+\Lambda_{3}.
\label{eq:WKB_nu}
\end{equation}
Here, $V_{0}''$ denotes the second derivative of the effective potential evaluated at its maximum, while $\Lambda_{2}$ and $\Lambda_{3}$ correspond to the second- and third-order WKB corrections. These corrections depend on derivatives of the effective potential up to sixth order, all evaluated at the location of the potential peak.

The second- and third-order WKB corrections, $\Lambda_{2}$ and $\Lambda_{3}$, take the standard form
\begin{widetext}
\begin{align}
\Lambda_{2} &=
\frac{1}{\left(2Q_{0}''\right)^{1/2}}
\left[
\frac{1}{8}\left(\frac{Q_{0}^{(4)}}{Q_{0}''}\right)\left(\frac{1}{4}+N^{2}\right)
-\frac{1}{288}\left(\frac{Q_{0}^{(3)}}{Q_{0}''}\right)^{2}\left(7+60N^{2}\right)
\right],
\\[0.3cm]
\Lambda_{3} &=
\frac{N}{2Q_{0}''}
\left[
\frac{5}{6912}\left(\frac{Q_{0}^{(3)}}{Q_{0}''}\right)^{4}\left(77+188N^{2}\right)
-\frac{1}{384}\left(\frac{ \left( Q_{0}^{(3)} \right)^2  Q_{0}^{(4)}}{Q_{0}''{}^{3}}\right)\left(51+100N^{2}\right)
\right.
\nonumber
\\
&\hspace{1.2cm}
\left.
+\frac{1}{2304}\left(\frac{Q_{0}^{(4)}}{Q_{0}''}\right)^{2}\left(67+68N^{2}\right)
+\frac{1}{288}\left(\frac{Q_{0}^{(3)}Q_{0}^{(5)}}{Q_{0}''{}^{2}}\right)\left(19+28N^{2}\right)
-\frac{1}{288}\left(\frac{Q_{0}^{(6)}}{Q_{0}''}\right)\left(5+4N^{2}\right)
\right],
\end{align}
\end{widetext}
where
\begin{equation}
N=\nu+\frac{1}{2}, \qquad
Q_{0}^{(n)}=\left.\frac{d^{n}Q}{dr_{*}^{n}}\right|_{r_{*}=r_{*}(r_{\max})},
\qquad Q\equiv \omega^{2}-V.
\end{equation}

The above expressions constitute the standard second- and third-order contributions in the WKB expansion. The full method was extended up to sixth order in Ref.~\cite{Iyer:1986np}, and has since been applied extensively to scattering problems around black holes (see, for instance, Ref.~\cite{GBWKB} and references therein).

The absorption cross section is given by
\begin{equation}
\sigma = \sum_{\ell=0}^{\infty} \sigma_{\ell},
\end{equation}
where each partial contribution is computed from the transmission coefficient through
\begin{equation}
\sigma_{\ell}
= \frac{\pi}{\omega^{2}} (2\ell + 1)\, |T(\omega)|^{2}.
\end{equation}

The WKB approximation has been extensively employed to evaluate the reflection and transmission coefficients, $R(\omega)$ and $T(\omega)$, and consequently the greybody factors in a wide variety of black-hole backgrounds.

The low–frequency regime cannot be accurately described by the WKB approximation, since the method requires the wavelength of the perturbation to be much smaller than the characteristic scale of variation of the potential barrier. When $\omega^{2}\ll V_{0}$ this condition is violated, and one must instead rely on the standard low–energy approximation, in which the transmission amplitude is exponentially suppressed. In this regime, the transmission coefficient is given by the well–known expression
\begin{equation}
T = e^{-\int_{z_{1}}^{z_{2}} dz\,\sqrt{V(z)-\omega^{2}}\,},
\label{eq:T_lowfreq_correct}
\end{equation}
where $z_{1}$ and $z_{2}$ are the classical turning points defined by $V(z)=\omega^{2}$. The reflection coefficient then follows as
\begin{equation}
R=\sqrt{\,1-e^{-2\int_{z_{1}}^{z_{2}} dz\,\sqrt{V(z)-\omega^{2}}\,}\,}.
\label{eq:R_lowfreq_correct}
\end{equation}

We emphasize that the low--frequency tunneling approximation,
Eqs.~(71)--(72), is applicable only when the effective potential
develops a genuine barrier, i.e. when two classical turning points
exist. In the present model, for sufficiently small frequencies
the equation $V(r)=\omega^2$ admits only a single root outside the
horizon, and therefore no classically forbidden region is formed.
In this regime the tunneling description is not applicable and the
reflection coefficient trivially approaches unity.
The low--energy formulas are thus evaluated only in the frequency
interval where two turning points are present. So, at very small frequencies the reflection coefficient approaches unity, meaning that nearly all the incoming flux is reflected by the potential barrier. As the frequency increases, the transmission grows gradually, and for values of $\omega$ comparable to or larger than the height of the barrier the reflection coefficient becomes negligible.

\begin{figure}[h]
\begin{center}
\includegraphics[width=0.35\textwidth]{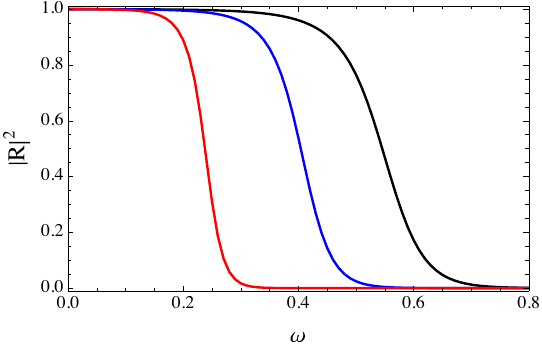}
\includegraphics[width=0.35\textwidth]{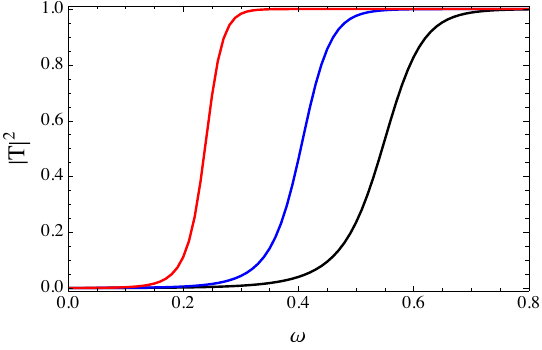}
\includegraphics[width=0.35\textwidth]{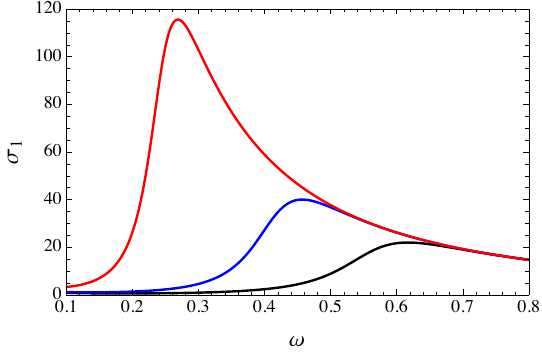}
\end{center}
\caption{Frequency dependence of the reflection coefficient $|R(\omega)|^{2}$ (top),
the transmission coefficient $|T(\omega)|^{2}$ (middle),
and the partial greybody factor $\sigma_{\ell}(\omega)$ (bottom), with $c=1$.
The different curves correspond to distinct values of the parameter $A$, $A=0.9$ (black curve), $A=1.0$ (blue curve), and $A=1.2$ (red curve)
while the angular momentum number $\ell$ is kept fixed ($\ell=1$).}
\label{coeff}
\end{figure}

Figure~\ref{coeff} displays the frequency dependence of the reflection coefficient 
$|R(\omega)|^{2}$ (top panel), the transmission coefficient 
$|T(\omega)|^{2}$ (middle panel), and the partial greybody factor 
$\sigma_{\ell}(\omega)$ (bottom panel), computed within the WKB 
approximation for different values of the parameter $A$, with $c=1$ and 
$\ell=1$. Increasing $A$ shifts the event horizon toward 
smaller radii and smooths the near-origin region, leading to systematic 
modifications of the effective potential shown in Fig.~\ref{fpot}. Although 
Fig.~\ref{fpot} is presented for a massive probe, its qualitative structure 
captures the essential features governing wave scattering. In 
particular, increasing $A$ the effective potential barrier becomes lower and 
narrower, thereby enhancing wave 
transmission at lower frequencies. This behavior is clearly reflected in 
Fig.~\ref{coeff}, where the transition from almost complete reflection at low 
frequencies to dominant transmission occurs at progressively smaller 
values of $\omega$ as $A$ increases. Correspondingly, the greybody factor 
exhibits a pronounced peak at intermediate frequencies, whose position 
shifts toward lower $\omega$ and whose magnitude is enhanced, signaling a more efficient absorption process.

\newpage

\section{Conclusions}
\label{FR}

In this work, we have investigated the dynamical response of asymptotically flat regular black holes supported by a phantom scalar field, focusing on scalar perturbations, quasinormal modes, and greybody factors. The regularizing scalar charge $A$ smooths out the central singularity while continuously deforming the Schwarzschild geometry, allowing us to assess how regularity effects manifest themselves in wave dynamics and scattering processes.

By analyzing the effective potential governing massive scalar perturbations, we have shown that increasing the scalar charge systematically lowers and broadens the potential barrier, shifting its maximum outward while preserving asymptotic flatness. These modifications have a direct impact on the quasinormal spectrum. Using WKB techniques beyond the eikonal limit, we have demonstrated that the presence of scalar hair reduces the oscillation frequencies and increases the damping times of the modes, in agreement with the weakened confinement induced by the regular geometry.  Furthermore, we have employed the Horowitz–Hubeny method and have shown that the results obtained from the WKB and Horowitz–Hubeny approaches exhibit excellent agreement in the regime where both methods are valid.

A central result of this work is the confirmation of an anomalous decay rate for massive scalar perturbations in this class of regular black holes. We have identified a critical value of the scalar field mass above which the hierarchy of damping times is inverted, so that modes with lower angular momentum become longer lived than those with higher multipole number. We derived analytical expressions for the critical mass and analyzed its dependence on the scalar charge and overtone number, and revealing how regularity effects shift the onset of the anomalous regime.

In addition, we have studied black-hole scattering by computing reflection and transmission coefficients and the associated greybody factors using WKB methods. Our analysis shows that the scalar charge leaves a clear imprint on the transmission probabilities, modifying both the height and width of the absorption window and thus affecting the frequency dependence of the greybody spectrum. These results provide a complementary characterization of regular black holes, linking quasinormal ringing and scattering properties within a unified wave-dynamical framework.

Overall, our findings indicate that regular black holes with scalar hair exhibit distinctive signatures in both their quasinormal mode spectrum and greybody factors. This reinforces the idea that wave dynamics offers a powerful probe of black-hole regularization mechanisms and scalar hair. Extensions of this work could include the analysis of other types of perturbations, different asymptotics, or the incorporation of observational constraints to further assess the phenomenological viability of regular black-hole models.

\appendix

\section{Full third-order WKB expressions for the QNFs and critical mass}
\label{appendixA}

In this appendix, we provide the complete analytical expressions for the quasinormal frequencies and the critical mass, derived using the WKB approximation at third order beyond the eikonal limit.

\begin{equation}
\label{A1}
\omega_{1m}=\frac{1}{A}\sqrt{\frac{\eta}{2}}\,,
\end{equation}
\begin{equation}
\label{A2}
\omega_{0}=-\frac{i (1+2n) \sqrt{\eta}}{2  \sqrt{2} A}\,, 
\end{equation}
\begin{eqnarray}
\notag \omega_{1} &=& \frac{1}{16 A}\sqrt{\frac{\eta}{2}} \Bigg[ 3 \pi  A^6 c^3 (30 n (n+1)-61)\\
\notag && -2 A^4 c^2 (30 n (n+1)-61)  \\
\notag && +3 \pi  A^2 c (2 n (n+1)-3)+8 \bar{m}^2 \left(9 A^6 c^2+A^2\right)  \\
\notag && -6 A^2 c \left(A^4 c^2 (30 n (n+1)-61)+2 n (n+1)-3\right) \\
&& \times \arctan  \left(3 A^2 c\right)-4 \left(n^2+n-1\right)\Bigg]\,, 
\end{eqnarray}
\begin{eqnarray} \label{omegas}
\notag \omega_2 &=& \frac{i (1+ 2n)}{128  \sqrt{2} A  } \eta^{3/2} \Bigg[   45 \pi  A^{10} c^5 \tilde{n}  
 -30 A^8 c^4 \tilde{n}+6 \pi  A^6 c^3 \tilde{n} \\
\notag && -188 A^4 c^2 \left(n^2+n-6\right)+3 \pi  A^2 c \left(n^2+n-15\right)+ \\
\notag && \bar{m}^2 \left(6480 A^{10} c^4+864 A^6 c^2+16 A^2\right)-6 A^2 c \big(15 A^8 c^4 \tilde{n}  \\
\notag && +2 A^4 c^2 \tilde{n}+n^2+n-15\big) \\
&& \times \arctan\left(3 A^2 c\right)-2 \left(n^2+n+3\right) \Bigg]\,,
\end{eqnarray}
where $\eta=2-3 \pi  c A^2 + 6 c A^2  \arctan \left( 3 c A^2\right)$, and $\tilde{n}=47n(n+1)-401$.
\newline

The critical mass $\bar{m}_c$ is determined by the condition $\omega_2 =0$, and is given by
\begin{eqnarray}
\label{mass}
\notag \bar{m}_c  &=& \frac{1}{4 A \sqrt{1+54 c^2 A^4 + 405 c^4 A^8}} \Bigg[  2 \big(15 A^8 c^4 \tilde{n} \\
\notag && +94 A^4 c^2 \left(n^2+n-6\right)+n^2+n+3\big) -  \\
\notag && 3 \pi  A^2 c \big(15 A^8 c^4 \tilde{n}+2 A^4 c^2 \tilde{n} \\
\notag && +n^2+n-15\big)+6 A^2 c \big(15 A^8 c^4 \tilde{n} +   \\
&& 2 A^4 c^2\tilde{n} +n^2+n-15\big) \tan ^{-1}\left(3 A^2 c\right)  \Bigg]^{1/2}.
\end{eqnarray}

\section{Near-horizon series expansions}
\label{appendixB}

In this appendix, we list the series expansions of the functions $s(x)$, $t(x)$ and $u(x)$ around the event horizon $x_+$, which are required for the implementation of the Horowitz--Hubeny method
\begin{eqnarray}
\notag s(x) &=& \sum_{n=0}^5 s_n (x-x_+)^n \,, \quad t(x) = \sum_{n=0}^5 t_n (x-x_+)^n \,,  \\
\notag u(x) &=& \sum_{n=0}^5 u_n (x-x_+)^n \,.
\end{eqnarray}
The expansion coefficients are given by
\begin{eqnarray}
\notag s_0 &=&  4 x_+^2 \left(A^2 x_+^2+1\right) \beta^2 \,, \\
\notag s_1 &=&  4 x_+ \left(5 A^2 x_+^2+3\right) \beta^2  \,,  \\
\notag s_2 &=&   4 \left(10 A^2 x_+^2+3\right) \beta^2  \,, \\
\notag s_3 &=&  \frac{4 }{x_+} \left(10 A^2 x_+^2+1\right)\beta^2 \,, \\
\notag s_4 &=& 20 A^2 \beta^2 \,,  \\
\notag s_5 &=& \frac{4  }{x_+} A^2  \beta^2\,,
\end{eqnarray}
\begin{eqnarray}
\notag t_0 &=& 4 x_+ \left(A^2 x_+^2+1\right) \left(x_+-2 i \omega \right) \beta^2  \,,   \\
\notag t_1 &=& 4 \omega ^2 \Bigg[x_+ \Big(3+A^2 x_+ \Big(7 x_+-2 i \Big( 2 \beta+5 \omega  \Big)  \Big) \Big)  \\
\notag &&   -2 i \left(2 \beta+3 \omega \right)\Bigg]+4 i \bar{m}^2 \Bigg[3 \bigg(\beta +2 \omega \bigg) \\
\notag && +x_+ \Big(A^2 x_+ \left(3 \beta+7 i x_++10 \omega \right)+3 i\Big)\Bigg] \,, \\
\notag t_2 &=&  \frac{4 \omega ^2}{x_+ } \Bigg[ x_+ \Big(3+2 A^2 x_+ \Big(9 x_+-i \Big(7 \beta+10 \omega \Big)\Big)\Big)  \\
\notag &&  -6 i \left(\beta+\omega \right)\Bigg]+ \frac{4 i \bar{m}^2 }{x_+ }  \Bigg [x_+ \Big(2 A^2 x_+ \Big(5  \\
\notag   && \times \left(\beta+2 \omega \right)+9 i x_+\Big)+3 i\Big)+4\beta   +6 \omega \Bigg]  \,, \\
\notag   t_3 &=&  \frac{4 \omega ^2}{x_+^2 } \Bigg[ 2 A^2 x_+^2 \left(11 x_+-i \left(9 \beta+10 \omega \right)\right) \\
\notag && -2 i \left(\beta+\omega \right)+x_+\Bigg]+\frac{4 i \bar{m}^2}{x_+^2 } \Bigg[  x_+ \Big(2 A^2 x_+  \\
\notag && \times \left(6 \beta+11 i x_++10 \omega \right)+i\Big)+\beta +2 \omega \Bigg]   \,,  \\
\notag  t_4 &=&  \frac{4 A^2}{x_+ }  \Bigg[ \omega ^2 \left(13 x_+-10 i \left(\beta+\omega \right)\right)  \\
\notag && +i \bar{m}^2 \left(6 \beta+13 i x_++10 \omega \right)\Bigg]  \,, \\
\notag t_5 &=& \frac{4 A^2}{x_+^2}  \Bigg[ \omega ^2 \left(3 x_+-2 i \left(\beta+\omega \right)\right) \\
\notag && +i \bar{m}^2 \left(\beta+3 i x_++2 \omega \right)\Bigg] \,,
\end{eqnarray}
 
\begin{eqnarray}
\notag u_0 &=& 0 \,, \\
\notag u_1 &=&   \frac{2 \bar{m}^2}{x_+} \Bigg[ 2 \omega  \left(3 \beta+5 \omega \right)+x_+  \Big(x_+ \Big(i A^2 x_+ \\
\notag && \times   \left(3 \beta+4 \omega \right)+2 A^2 \omega  \left(3 \beta+5 \omega \right)  \\
\notag &&  -2 \ell (\ell +1)\Big)+3 i \beta\Big)\Bigg]-\frac{4 \omega ^2 }{x_+} \Bigg[ 4 \omega  \Big(\beta+\omega \Big)  \\
\notag &&  + x_+ \Big(x_+ \Big(2 i A^2 x_+ \left(\beta+\omega \right)+  \\
\notag && 4 A^2 \omega  \left(\beta+\omega \right) - \ell ( \ell +1)\Big)+2 i \beta\Big)\Bigg] \\
\notag && -\frac{4 \bar{m}^4}{x_+} \left(A^2 x_+^2+1\right)  \,, \\
\notag u_2 &=&  \frac{2 x_+ \bar{m}^2 }{x_+^3}\Bigg[-2 x_+^2 \Big( -A^2 \omega \left(10 \beta+17 \omega \right)+\ell^2 +\ell \Big)\\
\notag &&  +i A^2 x_+^3 \left(11 \beta+16 \omega \right)-i x_+ \beta+   \\
\notag && 2 \omega  \left(4 \beta+7 \omega \right)\Bigg]-\frac{4 \omega ^2}{x_+^3} \Bigg[2 \left(\beta+\omega \right)  \\
\notag && \times  \left(7 A^2 x_+^3 \omega +4 i A^2 x_+^4+3 x_+ \omega \right)-\ell^2 x_+^3-\ell x_+^3\Bigg]  \\
\notag && -\frac{x_+ \bar{m}^4}{x_+^3} \left(13 A^2 x_+^2+5\right) \,, \\
\notag u_3 &=&  \frac{\bar{m}^2}{x_+^3} \Bigg[ 4 \omega  \left(\beta+2 \omega \right)+6 A^2 x_+^2 \Big(2 \omega  \Big(4 \beta  +7 \omega \Big)\\
\notag &&+i x_+ \left(5 \beta+8 \omega \right)\Big)\Bigg]-\frac{8 \omega ^2}{x_+^3} \Big(\beta +\omega \Big)\\
\notag &&  \left(\omega +3 A^2 x_+^2 \left(3 \omega +2 i x_+\right)\right)-\frac{\bar{m}^4 }{x_+^3}\left(15 A^2 x_+^2+1\right) \,,  \\
\notag u_4 &=&  \frac{2 \bar{m}^2  A^2 }{x_+^2} \Bigg[  2 \omega  \left(6 \beta+11 \omega \right)+i x_+ \Big(9 \beta \\
\notag && +16 \omega \Big)\Bigg]- \frac{ 8 \omega ^2 A^2 }{x_+^2}  \left(5 \omega +4 i x_+\right) \left(\beta+\omega \right)-\frac{7 \bar{m}^4  A^2 }{x_+^2} \,,  \\
\notag u_5 &=&  \frac{ 4 \bar{m}^2  A^2}{x_+^3} \left(\omega +i x_+\right) \left(\beta+2 \omega \right)-  \frac{8 \omega ^2A^2}{x_+^3} \left(\omega +i x_+\right)  \\
\notag && \times \left(\beta+\omega \right)- \frac{\bar{m}^4  A^2}{x_+^3} \,,
\end{eqnarray}
where $\beta^2=\omega^2 -\bar{m}^2.$

\clearpage
%\newpage

\begin{acknowledgments}

We thank the anonymous referee for valuable comments and suggestions. Y. V. acknowledges support by the Direcci\'on de Investigaci\'on y Desarrollo de la Universidad de La Serena, Grant No. PR25538511.

\end{acknowledgments}

\end{document}